\documentclass[journal,twoside,web]{IEEEtran}
\usepackage{cite}
\usepackage{amsmath,amssymb,amsfonts}
\usepackage{algorithmic}
\usepackage{graphicx}
\usepackage{textcomp}
\usepackage{bm}
\usepackage{siunitx}							%correct display of SI units
\usepackage{tabularx}							%for centered tables
\newcolumntype{C}[1]{>{\centering\arraybackslash}m{#1}}	%new column type for centered columns
\usepackage{epsfig}
\usepackage{scalerel}
\usepackage{tikz}
\usetikzlibrary{svg.path}

\definecolor{orcidlogocol}{HTML}{A6CE39}
\tikzset{
  orcidlogo/.pic={
    \fill[orcidlogocol] svg{M256,128c0,70.7-57.3,128-128,128C57.3,256,0,198.7,0,128C0,57.3,57.3,0,128,0C198.7,0,256,57.3,256,128z};
    \fill[white] svg{M86.3,186.2H70.9V79.1h15.4v48.4V186.2z}
                 svg{M108.9,79.1h41.6c39.6,0,57,28.3,57,53.6c0,27.5-21.5,53.6-56.8,53.6h-41.8V79.1z M124.3,172.4h24.5c34.9,0,42.9-26.5,42.9-39.7c0-21.5-13.7-39.7-43.7-39.7h-23.7V172.4z}
                 svg{M88.7,56.8c0,5.5-4.5,10.1-10.1,10.1c-5.6,0-10.1-4.6-10.1-10.1c0-5.6,4.5-10.1,10.1-10.1C84.2,46.7,88.7,51.3,88.7,56.8z};
  }
}

\newcommand\orcidicon[1]{\href{https://orcid.org/#1}{\mbox{\scalerel*{
\begin{tikzpicture}[yscale=-1,transform shape]
\pic{orcidlogo};
\end{tikzpicture}
}{|}}}}

\usepackage{hyperref}							%required for urls with underscore

\def\BibTeX{{\rm B\kern-.05em{\sc i\kern-.025em b}\kern-.08em
    T\kern-.1667em\lower.7ex\hbox{E}\kern-.125emX}}
\begin{document}
\title{Influence of Different Surface Morphologies on the Performance of High Voltage, Low Resistance Diamond Schottky Diodes}
\author{Philipp Reinke \orcidicon{0000-0002-3644-2738}, Fouad Benkhelifa \orcidicon{0000-0002-6255-117X}, Lutz Kirste \orcidicon{0000-0002-5274-2650}, Heiko Czap \orcidicon{0000-0003-4454-1215}, Lucas Pinti \orcidicon{0000-0002-4319-0954}, Verena Zürbig \orcidicon{0000-0002-4862-1303}, Volker Cimalla \orcidicon{0000-0001-8531-1892}, Christoph Nebel \orcidicon{0000-0002-7976-7030}, Oliver Ambacher \orcidicon{0000-0001-5193-9016}
\thanks{Manuscript received February 6, 2020; revised April 8, 2020; accepted April 20, 2020.}
\thanks{This work was supported by the project DiaLe from the Fraunhofer-Gesellschaft e.V. (Corresponding author: Philipp Reinke) }
\thanks{The authors P. Reinke, F. Benkhelifa, L. Kirste, H. Czap, V. Cimalla, C. Nebel and O. Ambacher are with the Fraunhofer IAF, Fraunhofer Institute for Applied Solid State Physics, 79108 Freiburg, Germany (e-mail: philipp.reinke@iaf.fraunhofer.de).}
\thanks{The author V. Zuerbig is with the Infineon Technologies AG, 85579 Neubiberg, Germany.}
\thanks{The author L. Pinti is with the Sonova Holding AG, 8712 Stäfa, Switzerland.}
\thanks{\copyright 2020 IEEE.Personal use of this material is permitted. Permission from IEEE must be obtained for all other uses, in any current or future media, including reprinting/republishing this material for advertising or promotional purposes,creating new collective works, for resale or redistribution to servers or lists, or reuse of any copyrighted component of this work in other works.}
\thanks{Digital Object Identifier DOI \href{https://doi.org/10.1109/TED.2020.2989733}{10.1109/TED.2020.2989733}}}

\maketitle

\begin{abstract}
Vertical diamond Schottky diodes with blocking voltages V\textsubscript{BD} \textgreater{} 2.4 kV and on-resistances R\textsubscript{On} \textless{} 400 m$\Omega$cm\textsuperscript{2} were fabricated on homoepitaxially grown diamond layers with different surface morphologies. The morphology (smooth as-grown, hillock-rich, polished) influences the Schottky barrier, the carrier transport properties, and consequently the device performance. The smooth as-grown sample exhibited a low reverse current density J\textsubscript{Rev} \textless{} 10\textsuperscript{-4} A/cm\textsuperscript{2} for reverse voltages up to 2.2 kV. The hillock-rich sample blocked similar voltages with a slight increase in the reverse current density (J\textsubscript{Rev} \textless{} 10\textsuperscript{-3} A/cm\textsuperscript{2}). The calculated 1D-breakdown field, however, was reduced by 30 \%, indicating a field enhancement induced by the inhomogeneous surface. The polished sample demonstrated a similar breakdown voltage and reverse current density as the smooth as-grown sample, suggesting that a polished surface can be suitable for device fabrication. However, a statistical analysis of several diodes of each sample showed the importance of the substrate quality: A high density of defects both reduces the feasible device area and increases the reverse current density. In forward direction, the hillock-rich sample exhibited a secondary Schottky barrier, which could be fitted with a modified thermionic emission model employing the Lambert W-function. Both polished and smooth sample showed nearly ideal thermionic emission with ideality factors 1.08 and 1.03, respectively. Compared with literature, all three diodes exhibit an improved Baliga Figure of Merit for diamond Schottky diodes with V\textsubscript{BD} \textgreater{} 2 kV.
\end{abstract}


\section{Introduction}
\label{sec:introduction}
\IEEEPARstart{W}{ith} its unique electrical and thermal properties, diamond possesses a high potential for the use in power electronic devices. In particular, its high thermal conductivity of $\SI{2000}{W/(m.K)}$, the hole mobility of up to $\SI{3800}{cm^2/(V.s)}$, and the theoretical dielectric strength of $\SI{10}{MV/cm}$ \cite{Nazare.2001}\cite{Isberg.09.2002}\cite{Umezawa.05.2018}, make diamond a viable choice for addressing key issues (e.g. switching losses, heat dissipation, high breakdown voltages) of power electronic devices.

Schottky barrier diodes (SBD), either vertical (VSBD), pseudo vertical (pVSBD) or planar (plSBD), are commonly used to assess the feasibility of diamond as a material for power electronic applications \cite{Umezawa.05.2018}. Although it has been shown that individual diodes are capable of blocking high voltages ($V_{\text{BD}}>\SI{10}{kV}$ \cite{Volpe.08.2010}), withstanding high electric fields ($E_{\text{Max}}=\SI{9.5}{MV/cm}$ \cite{Volpe.11.2010}) or of transporting high currents ($I_{\text{Max}}>\SI{20}{A}$ \cite{Bormashov.05.2017}), there has been no reported research on diodes exhibiting both high breakdown voltages $V_{\text{BD}}$  and high currents $I_{\text{Max}}$ to this date. Umezawa \cite{Umezawa.05.2018} identifies missing edge terminations, high reverse current through electrically relevant defects, and a non-optimized device design as key causes for their performance far below the theoretical expectations.

Additionally, both surface termination \cite{Volpe.08.2010}\cite{Teraji.06.2009} and surface morphology play an important role in the performance of diamond SBDs. Umezawa \textit{et al.} \cite{Umezawa.02.2007} identified nonepitaxial crystallites on the surface as a cause for increased leakage currents. Furthermore, Teraji at. al. \cite{Teraji.05.2012} analyzed the influence of mid-gap states on the reverse current $I_{\text{Rev}}$  but did not observe any influence of the surface roughness on $I_{\text{Rev}}$.

However, the analysis of SBDs on polished epitaxial surfaces has not been investigated. In this paper, we therefore present the performance of high voltage ($V_{\text{BD}}>\SI{2.4}{kV}$ ) vertical Schottky barrier diodes with low specific on-resistances ($R_{\text{on}}A = (\num{300} - \num{400}) \: \si{m\ohm.cm^2}$) fabricated on homoepitaxial diamond with different surface morphologies (Hillocks on surfaces, smooth as-grown, polished). The diodes are compared with published diamond diodes exhibiting similar blocking voltages (Twitchen \textit{et al.} \cite{Twitchen.05.2004}) and exhibit an approx. 7-time increase in the Baliga Figure of Merit $\text{BFOM} = V_{\text{BD}}^2/(R_{\text{on}}A)$. Recently, diamond diodes fabricated with the same methods as presented in this work were used to demonstrate the application in a non-isolated buck converter \cite{Reiner.05.2019}.

\section{Experimental details} 

\begin{figure}[!t]
    \centerline{\includegraphics[width=\columnwidth]{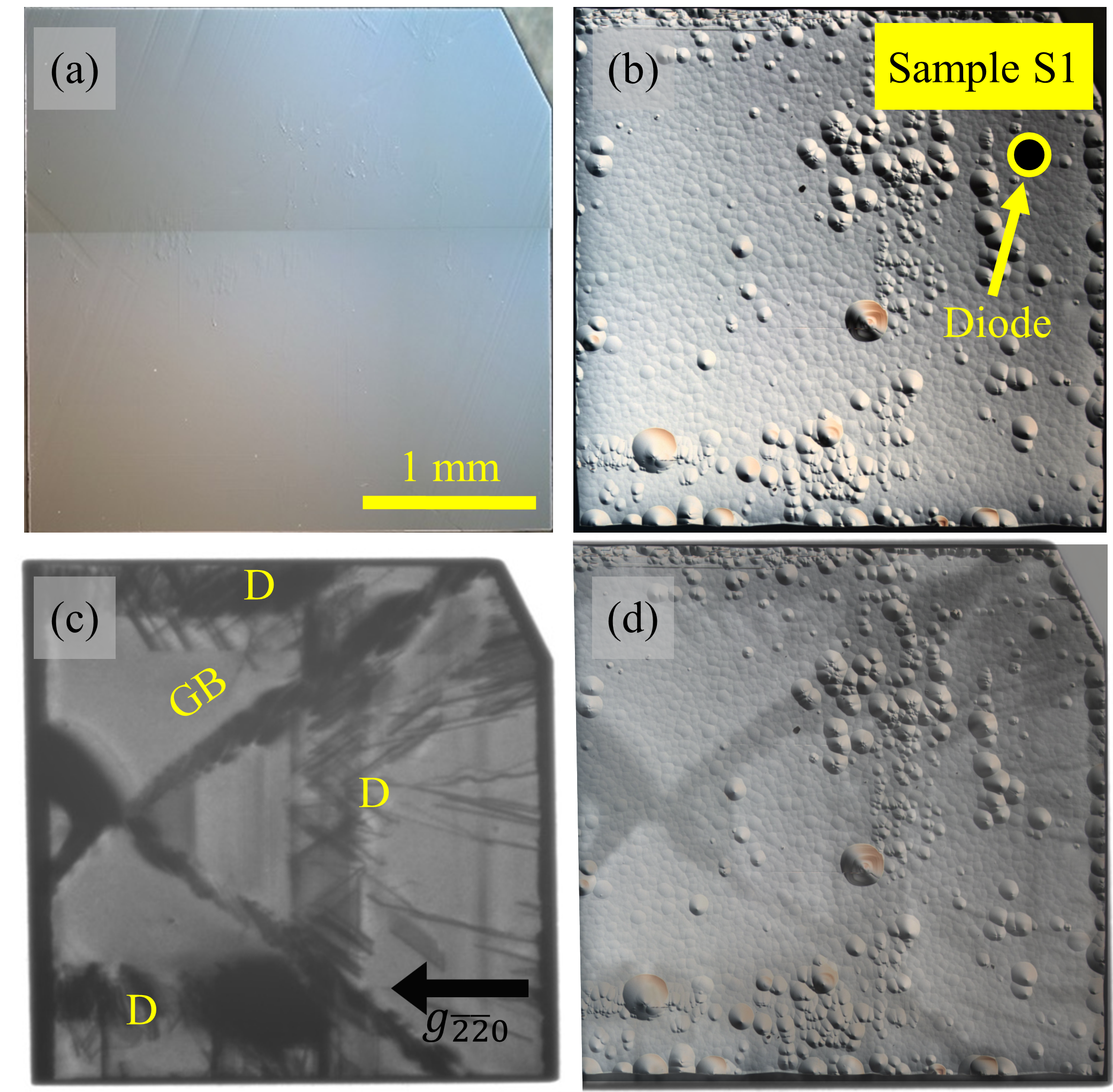}}
    \caption{Microscopic image of the sample S1 (a) before and (b) after growth of a 28 µm intrinsic diamond layer. The position of the diode is marked in yellow. In the X-ray topography image (c), defects (D) and growth sector boundaries (GB) can be identified. Image (d) shows an overlay of (b) and (c).}
    \label{fig1}
\end{figure}

The vertical diamond Schottky diodes were realized using 17 µm to 28 µm thick, unintentionally doped ($N_A < \SI{{}e15}{cm^{-3}}$) homoepitaxial diamond i-layers, which were grown on thick (300 µm) highly boron doped ($N_A \approx \SI{2e20}{cm^{-3}}$, type IIb), commercial HPHT substrates. The i-layers were grown in a home-made microwave plasma-enhanced chemical vapor deposition (MWPECVD) system (for details of the deposition system see \cite{Funer.03.1998} and \cite{Widmann.04.2016}). Prior to growth, the polished substrates were acid cleaned in a 3:1 mixture of sulfuric acid and nitric acid at $\SI{250}{\degreeCelsius}$ for 90 minutes, with a subsequent methanol rinse to remove any particles. In order to minimize the influence of any polishing damage, a H\textsubscript{2}-plasma etch was conducted for 5 to 40 minutes directly before the initiation of the growth. The growth conditions were optimized to reduce residual boron doping and to create a surfaces with a reduced roughness.  They were as follows: A gas mixture with a CH\textsubscript{4}/H\textsubscript{2} ratio of 4 \% (Sample S1) and 3 \% (Samples S2 and S3) with an additional 0.15 \% O\textsubscript{2}/H\textsubscript{2} \cite{Teraji.2015} was used, at a process pressure of 200 mbar. The microwave power was adjusted in the range of 2.1 kW to 2.3 kW to stabilize a substrate temperature of roughly $\SI{800}{\degreeCelsius}$. During growth, the samples were rotated to minimize any anisotropic effects arising from the direction of the gas flow.

After growth, the samples were cleaned in the mixture of sulfuric acid and nitric acid once more to remove any graphitic parts of the backside and to enhance the oxygen termination of the surface. Apart from sample S3, which was mechanically polished on an iron plate to flatten the hillocks observed after growth, all samples were used as-grown. To reduce the sub-surface polishing damage, the polished sample S3 was etched for 10 minutes in a 2.2 kW H\textsubscript{2} plasma at $\SI{800}{\degreeCelsius}$ in the MWPECVD prior to contact fabrication. Microscopic images of the layers used for device fabrication are shown in Fig. \ref{fig1} (b), Fig. \ref{fig2} (a) and Fig. \ref{fig3} (b) for the samples S1, S2 and S3, respectively.

The ohmic contacts were realized with a stack of TiPtAu deposited by electron-beam evaporation on the backside of the highly boron doped substrates. A subsequent rapid thermal annealing in N\textsubscript{2} atmosphere for 60 seconds at $\SI{850}{\degreeCelsius}$ yielded an ohmic contact with a contact resistivity of $\rho = \SI{1e-5}{\ohm.cm^2}$. Before evaporation of the Schottky metal, a short oxygen plasma ashing was used to enhance the oxygen termination of the surface. Using standard lithography techniques, several Schottky diodes with diameters d = \{100, 200, 300\} µm were fabricated on each of the samples. Electron-beam evaporated titanium was used as a Schottky metal with platinum as a diffusion stop layer and gold as a capping layer. Exemplarily, Fig.  \ref{fig3} (c) shows the diodes of the sample S3.

The electrical properties of the vertical Schottky diodes were analyzed by capacitance voltage measurements (CV) and current voltage measurements (IV) under forward and reverse bias. To avoid arcing at high reverse voltages, the diodes were immersed in Fluorinert during reverse IV measurement.

\section{Results and discussion}

\subsection{Sample morphology}

\begin{figure}[!t]
    \centerline{\includegraphics[width=\columnwidth]{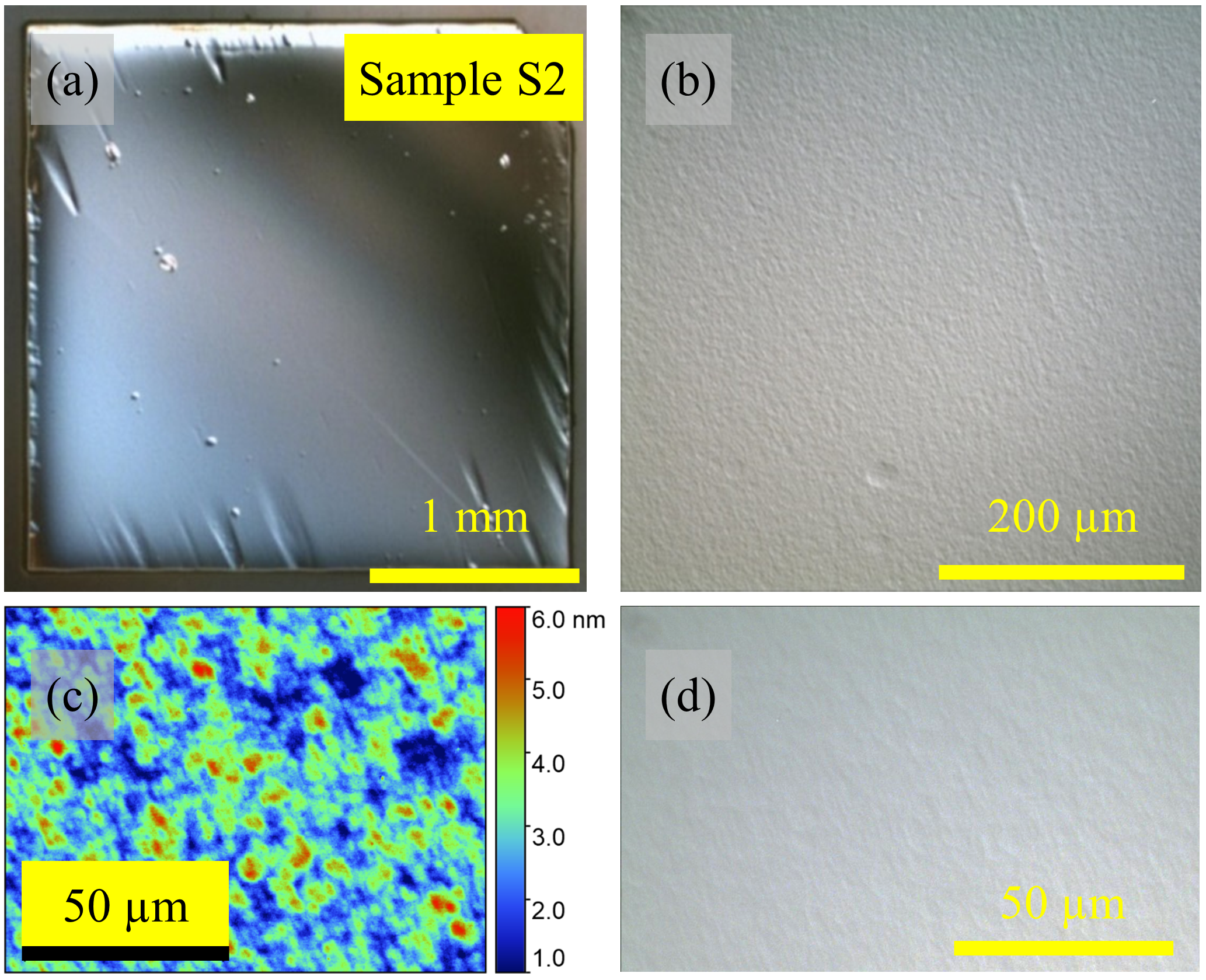}}
    \caption{(a) Overview, (b) and (d) detailed microscopic images, and (c) WLI surface roughness of sample S2 after growth of a 17 µm thick intrinsic diamond layer with a reduced CH\textsubscript{4}/H\textsubscript{2} ratio of 3 \%.}
    \label{fig2}
\end{figure}

All three diodes discussed in this work were chosen to have a similar maximum blocking voltage of $V_{\text{BD}} \approx \SI{2.5}{kV}$ despite being processed on diamond samples with different surface morphologies.

In Fig. \ref{fig1}, a microscopic image of the hillock rich sample S1 is presented (a) before and (b) after growth. On the right half of the sample, a higher density of growth hillocks is observed. These larger hillocks presumably arise from defects which are present in the substrate \cite{Tallaire.01.2008}: The overlay of the X-ray topography of S1 [Fig. \ref{fig1} (c)] with the microscopic image in Fig. \ref{fig1} (d) suggests a correlation of the growth hillocks with the defects present in the substrate. A detailed analysis of the hillock formation is outside the scope of this paper. The area in the center of the left half of the sample is characterized by a rough surface (RMS roughness 12 nm) with less hillocks. The relatively high roughness is believed to originate from the high CH\textsubscript{4}/H\textsubscript{2} ratio of 4 \%. The diode on S1, which is analyzed in this paper, is situated in the rough but hillock free region.

Since a rough surface can have negative effects on the electrical characteristics such as field enhancement or secondary Schottky barriers, sample S2 was grown with a reduced CH\textsubscript{4}/H\textsubscript{2} ratio of 3 \%. The microscopic images in Fig. \ref{fig2} (a) and (b) show a significantly smoother surface compared to the sample S1, with an RMS roughness \textless{} 0.8 nm [Fig. \ref{fig2} (c), measured with a white light interferometer (WLI)]. Furthermore, no prominent growth hillocks are observable over the entire sample, indicating a higher substrate quality with less defects.

\begin{figure}[!t]
    \centerline{\includegraphics[width=\columnwidth]{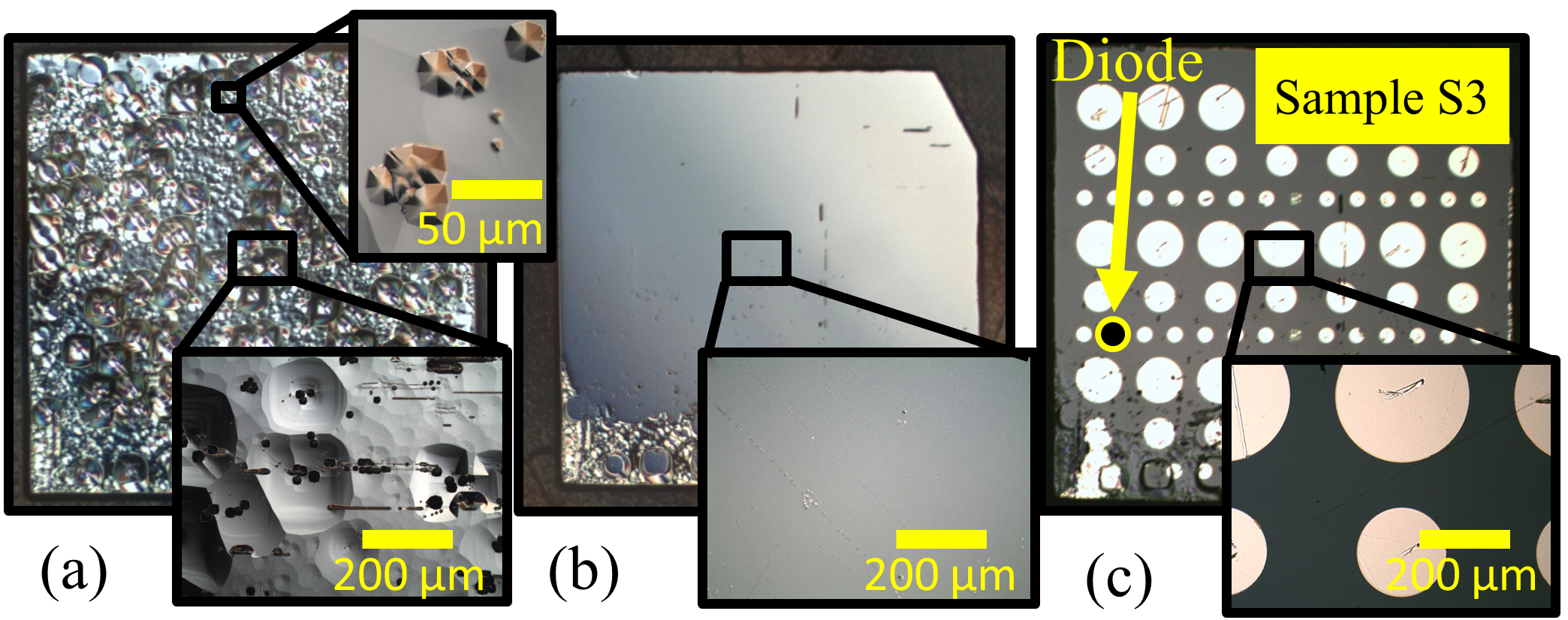}}
    \caption{Microscopic images of sample S3 (a) after growth of a 26 µm intrinsic diamond layer, (b) after polishing and a subsequent H\textsubscript{2} plasma etch and (c) after processing of the Schottky contacts. The diode being tested is marked.}
    \label{fig3}
\end{figure}

Although the third sample S3 was grown under the same conditions as S2, a surface with a high density of hillocks and etch pits was observed (Fig. \ref{fig3} (a), RMS roughness \textgreater{} 50 nm). A low quality substrate with a high density of dislocations is assumed to be the cause for these undesired features: The substrate’s dislocations penetrate into the epitaxial layer during growth and become sites for preferential etching and hillock growth \cite{Tallaire.01.2008}. These defects are assumed to be “killer defects” \cite{Ohmagari.09.2011}\cite{Akashi.06.2018}\cite{Kato.08.2015} for the Schottky diode and can cause a premature breakdown. To reduce undesired effects of the surface morphology on the diode performance, ca. 3 µm were removed by polishing of the top surface [Fig. \ref{fig3} (b)], leaving a surface with a RMS roughness of 1.5 nm. Note that due to a slight misorientation during polishing, the surface is inclined by 0.13°, as can be seen by the non-polished area in the lower left corner of the sample.

\subsection{Electrical measurements}

\subsubsection{CV-characteristics}

\begin{figure}[!t]
    \centerline{\includegraphics[width=\columnwidth]{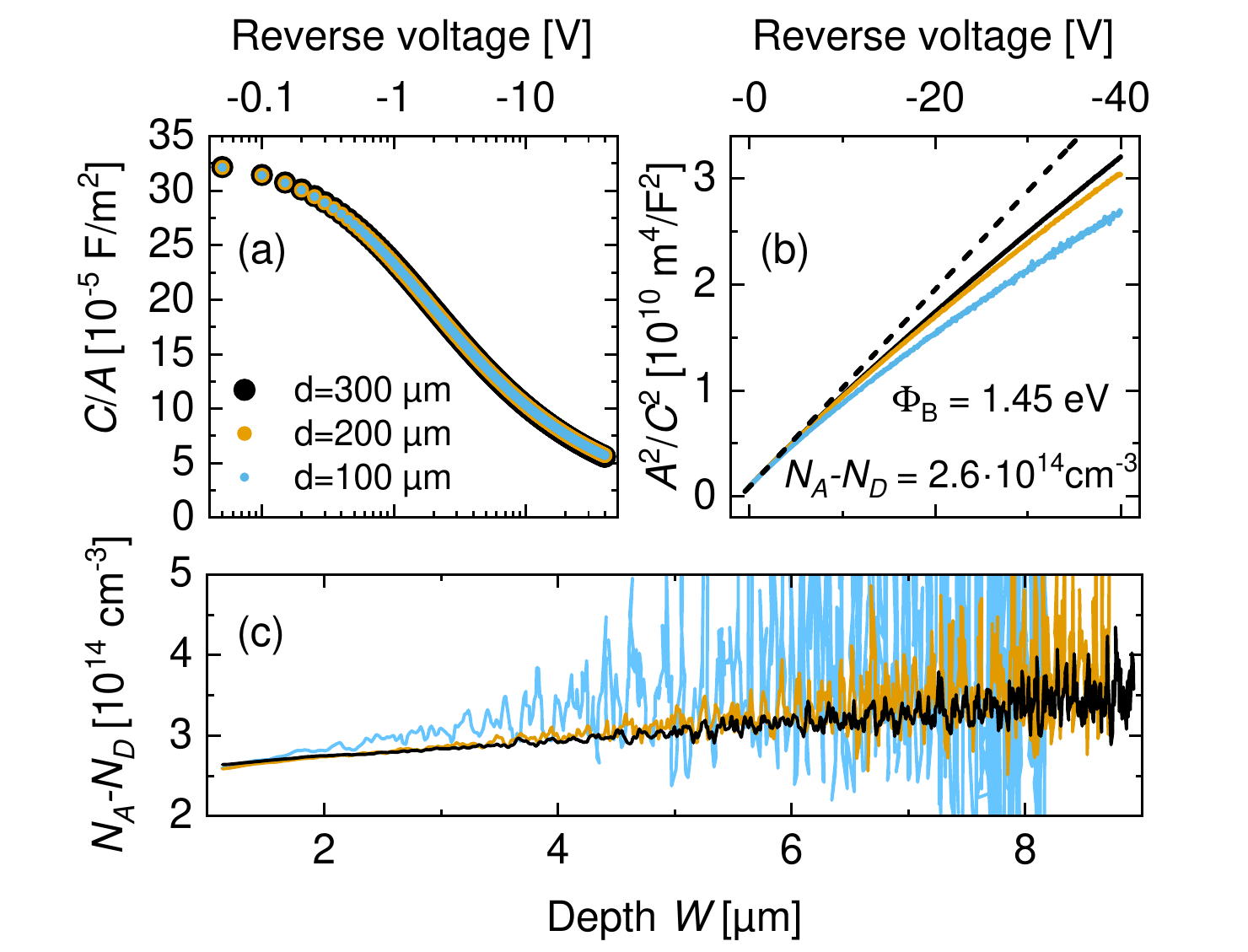}}
    \caption{(a) Area specific capacitance $C/A$ of three differently sized diodes of sample S2 and (b) extracted $A^2/C^2$ plot to estimate the doping density $N_A-N_D$. (c) Shows $N_A-N_D$ versus depth.}
    \label{fig4}
\end{figure}

Fig. \ref{fig4} (a) shows the capacitance $C$ per area $A$ from sample S2 for three differently sized diodes. There is no variance in $C/A$ in between the diodes with different diameters. Fig. \ref{fig4} (b) plots the calculated $A^2/C^2$ versus the reverse voltage. The net doping concentration $N_A - N_D$ is calculated with a linear fit between 0 V and 1 V using $N_A - N_D = \frac{2}{q \epsilon_0 \epsilon_d} \times \left( -\frac{\text{d}(A^2/C^2)}{\text{d}V} \right)^{-1}$,
%\begin{equation}
%	N_A - N_D = \frac{2}{q \epsilon_0 \epsilon_d} \times \left( -\odv{(A^2/C^2)}{V} \right)^{-1}
%	\label{eq_NAND}
%\end{equation}
where $q$ denotes the elementary charge, $\epsilon_0$ the vacuum permittivity, and $\epsilon_d$ the diamond’s permittivity \cite{Sze.2007}. This yields $N_A - N_D = \SI{2.6e14}{cm^{-3}}$ for all three diodes of sample S2. With $N_A - N_D = \SI{6.8e14}{cm^{-3}}$ and $N_A - N_D = \SI{2.0e14}{cm^{-3}}$, respectively, Sample S1 and sample S3 exhibit similar net doping concentrations.

From the built-in voltage $\psi_{\text{Bi}}$ given by the x-axis intercept of the linear fit of the $A^2/C^2$ data, the Schottky barrier height $\varphi_B$ is calculated by $\varphi_B = \psi_{\text{Bi}} + k_B T + \varphi_n$ \cite{Sze.2007}. Here, $\varphi_n = (E_F - E_V)/q \approx \SI{0.36}{eV}$ \cite{Nebel.2017} is the energy difference between valence band and Fermi level, with the temperature T and Boltzmann constant $k_B$. The NextNano software was employed for the simulation of $\varphi_n$ using the corresponding net doping concentration. Both sample S1 and sample S2 have similar $\varphi_B = \SI[separate-uncertainty = true]{1.45(1)}{eV}$, whereas the polished sample S3 exhibited a much larger barrier height of $\varphi_B = \SI[separate-uncertainty = true]{1.80(5)}{eV}$ (averaged over six diodes each).

With a simple plate capacitor model to estimate the depletion depth $W=\epsilon_0 \epsilon_d A/C$, the differential slope of $A^2/C^2$ can be used to plot $N_A-N_D$ versus $W$. As can be seen in Fig. \ref{fig4} (c), the net doping concentration is nearly constant over the probed depth. Due to their smaller absolute capacitance values, the smaller diodes are affected more heavily by signal noise.

The $N_A-N_D$ value as measured with CV was confirmed by cathodoluminescence (CL) measurements \cite{Barjon.02.2011} on other layers grown under the same condition, which yielded the same net doping concentration for both measurement types. The analysis of the CL data is not the scope of this paper and will be discussed elsewhere.

\subsubsection{IV-characteristics}

\begin{figure}[!t]
    \centerline{\includegraphics[width=\columnwidth]{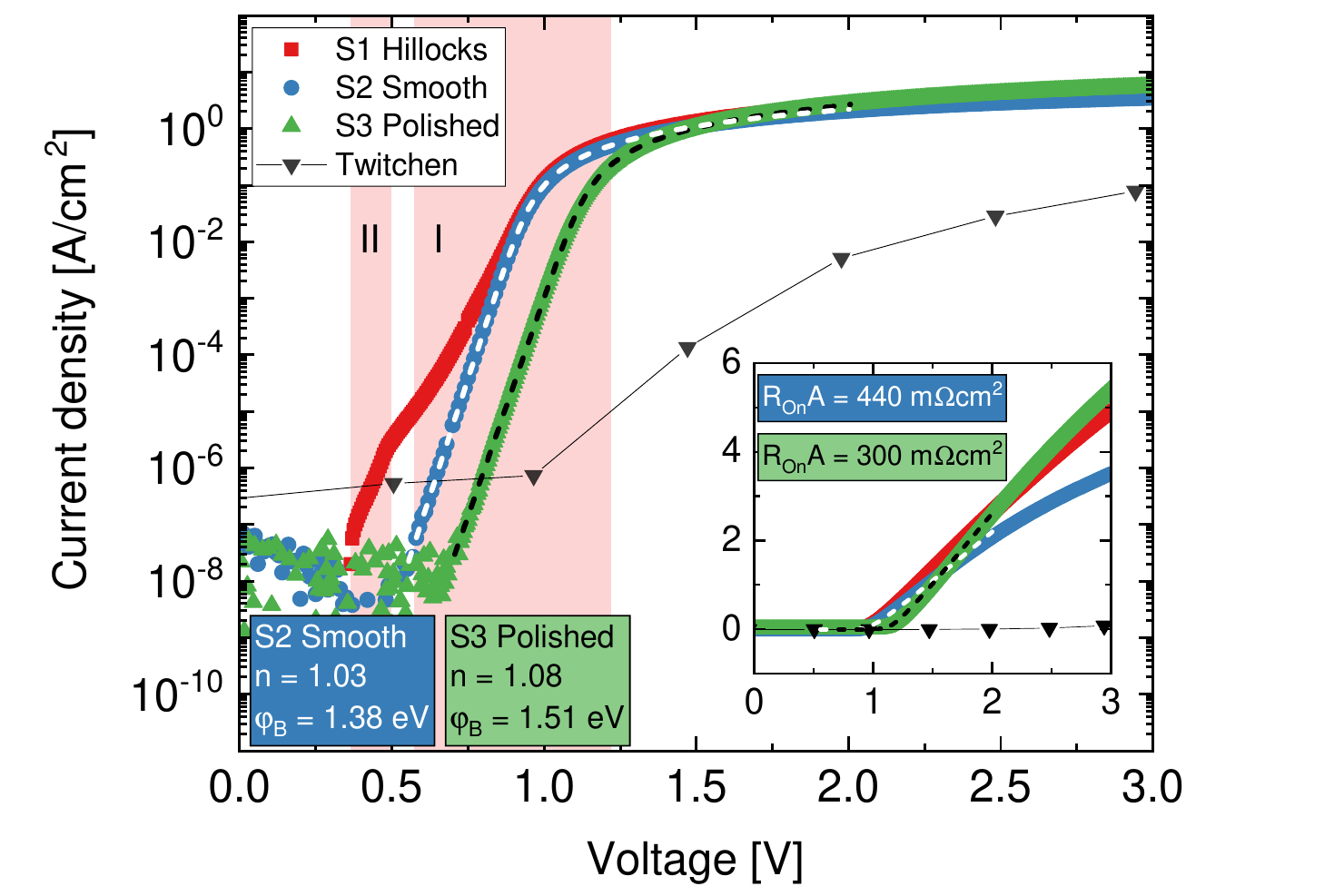}}
    \caption{Forward IV curves of the high voltage blocking diodes of the three samples S1, S2 and S3 in comparison with data from \cite{Twitchen.05.2004}. The inset shows the IV data in linear scale. The dashed lines represent fits to formula (3) which yield $n$, $\varphi_B$ and $R_{\text{on}}A$. The diode on the rough sample S1 exhibits two distinct regions (labeled I and II) of the current transport, see Fig. \ref{fig6} for details.}
    \label{fig5}
\end{figure}

To study the electrical properties of the Schottky contact, the forward IV characteristics are analyzed. Fig. \ref{fig5} compares the IV characteristics of the three high voltage blocking diodes of the samples S1, S2 and S3 with the data reported by Twitchen \textit{et al.} \cite{Twitchen.05.2004} for a diamond Schottky diode with a similar breakdown voltage. Both diodes on sample S2 (smooth as grown) and sample S3 (smooth after polishing) show a very good agreement with the thermionic emission (TEM) current transport model
\begin{equation}
    J_{\text{TEM}} = A^{\ast} T^2 e^{-\frac{q \varphi_B}{k_B T}} \left( e^\frac{q(V - J R_{on} A)}{n k_B T} - 1 \right)
   \label{eq_TEM}
\end{equation}
represented by the dashed lines in Fig. \ref{fig5}. The model includes an ohmic term $R_{\text{on}}A$ to account for the series resistance of the i-layer. $A^{\ast} = 4 \pi m^{\ast} q k_B^2 / h^3 = \SI{90}{A.cm^{-2}.K^{-2}}$ is the Richardson constant for diamond \cite{Blank.08.2015}. With an ideality factor of $n=\num{1.03}$ and $n=\num{1.08}$ for the smooth as-grown and polished sample, respectively, the current transport is nearly purely thermionic. The barrier height, on the other hand, is noticeably higher on the polished sample S3 ($\varphi_B = \SI{1.51}{eV}$) than on the smooth as-grown sample S2 ($\varphi_B = \SI{1.38}{eV}$). This might be a consequence of the additional H\textsubscript{2} etch step on S3 after polishing.

The inset of Fig. \ref{fig5} shows a lin-lin plot of the same IV data. The diode on the polished sample S3 exhibits a reduced on-resistance of $R_{\text{on}}A = \SI{300}{m\ohm.cm^2}$ compared to $R_{\text{on}}A = \SI{440}{m\ohm.cm^2}$ of the diode on the smooth as-grown sample S2. Considering the different layer thicknesses $d_{\text{i-layer}}$, the series resistivity $\rho = R_{\text{on}}A / d_{\text{i-layer}}$ of the diode on the smooth as-grown sample S2 is roughly twice the series resistivity of the diode on the polished sample S3. This is surprising, since their quite similar doping level would only account for a difference of 5 \% in the resistivity. Future work will include the investigation of the temperature dependence of $R_{\text{on}}A$ as well as Hall measurements to further examine the reduced series resistance of the polished sample.

When comparing the IV data of the three samples from this work with the data reported by Twitchen \cite{Twitchen.05.2004}, the high on-resistance of the latter is evident. From the IV slope at higher voltages ($U > \SI{6}{V}$), an $R_{\text{on}}A \approx \SI{2000}{m\ohm.cm^2}$ is estimated for Twitchen’s diode.

\begin{figure}[!t]
    \centerline{\includegraphics[width=\columnwidth]{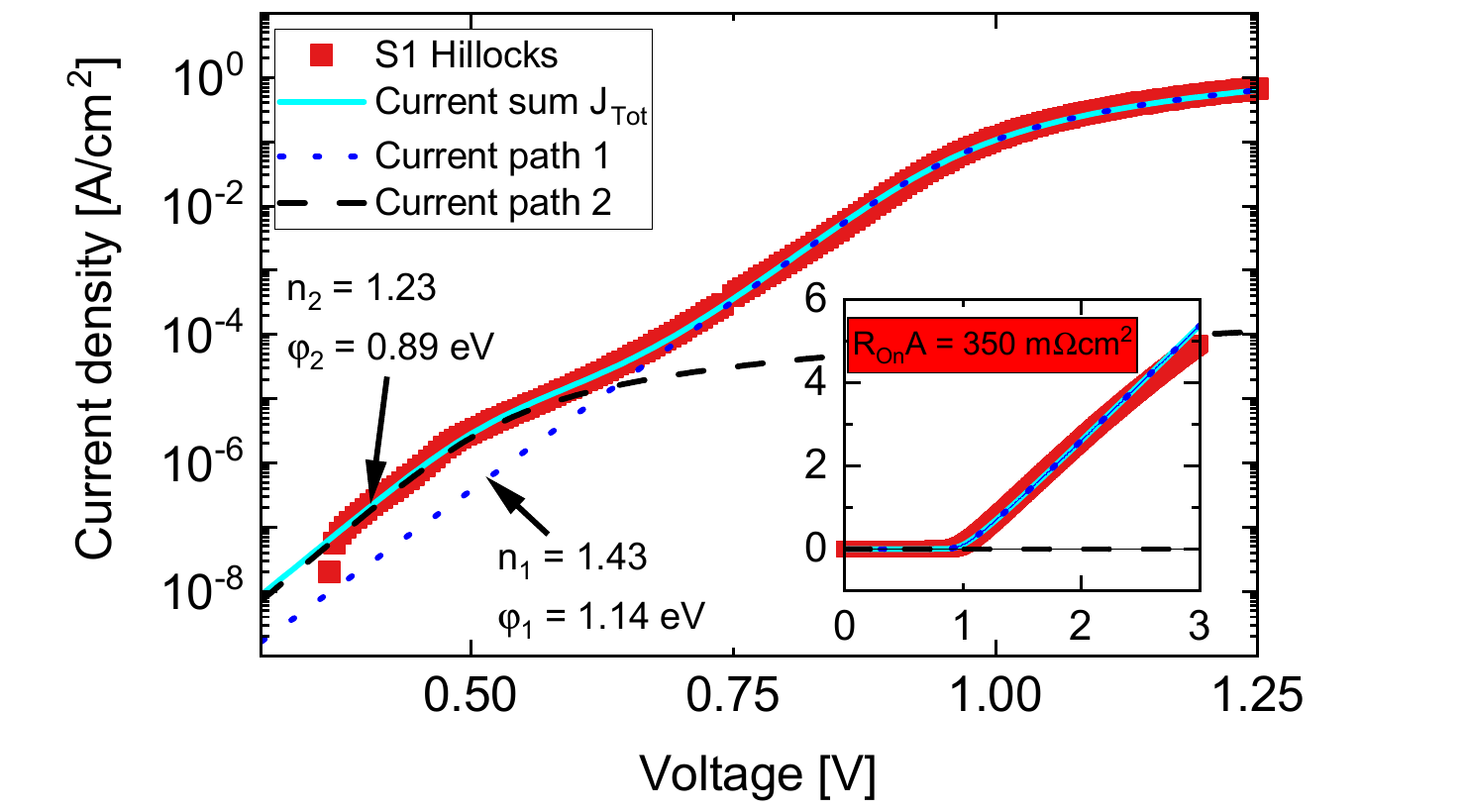}}
    \caption{Detail of the forward IV curve from sample S1 (linear scale in inset). The data is fitted to (\ref{eq_LambertW}). The current transport in the low voltage range is dominated by a secondary barrier $\varphi_{2} = \SI{0.89}{eV}$. Details see text.}
    \label{fig6}
\end{figure}

The IV characteristic of the diode on the rougher sample S1 exhibits two different slopes in the exponential region. In Fig. \ref{fig5}, the two regions with differing slopes are marked with light red backgrounds (labeled as I and II), with a detailed view shown in Fig. \ref{fig6}. As discussed in \cite{Defives.03.1999}\cite{Ma.01.2006} and \cite{Umezawa.09.2009} for SiC and diamond Schottky diodes, respectively, this can have its origin in localized patches with a secondary, lower Schottky barrier height $\varphi_{B\text{-low}}$. Defives \textit{et al.} \cite{Defives.03.1999} used an iterative approach to fit formula (\ref{eq_TEM}) to the two different regions, using the theoretically expected series resistance to extract the barrier heights and ideality factors. Note that due to the implicit form of (\ref{eq_TEM}), this iterative approach does not allow for a combined IV model for the two regions.

As was shown by Banwell \textit{et al.} \cite{Banwell.02.2000} and later adapted by Jung \textit{et al.} \cite{Jung.11.2009}, an explicit analytical solution exists for (\ref{eq_TEM}) by using the Lambert W-function. As was recently analyzed by Olikh \cite{Olikh.07.2015}, extracting the Schottky diode parameters by the Lambert W-function reduces both the determination error and the number of accuracy influencing factors. Ortiz-Conde \textit{et al.} \cite{OrtizConde.06.2012} used the explicit nature of the Lambert W-function to model the multiexponential behavior of solar cells.

A similar approach will be used in the following to assess the Schottky parameters of the diode on the “hillock sample” S1. The total current through the diode $I_{\text{Tot}}$ is modelled by the sum of the current through $N$ branches and given by
\begin{equation}
    I_{\text{Tot}}(V) = \sum_{i=1}^N \frac{n_i k_B T}{q R_{\text{on},i}} \text{W}_0 \left( \frac{I_{0,i} R_{\text{on},i} q}{n_i k_B T} e^{\frac{q (V + I_{0,i} R_{\text{on},i} )}{n_i k_B T} } \right) - I_{0,i}.
   \label{eq_LambertW}
\end{equation}
For each branch $i$ the following parameters are used: The saturation current $I_{0,i} = A_i A^{\ast} T^2 \exp{-\beta \varphi_{B,i}}$, the active Schottky contact area $A_i$, the ideality factor $n_i$, the series resistance $R_{\text{on},i}$, and the Schottky barrier height $\varphi_{B,i}$. $\text{W}_0 (x)$ is the Lambert W-function \cite{Corless.12.1996} and $\beta = q/(k_B T)$.

Formula (\ref{eq_LambertW}) with $N=\num{2}$ was used to fit the IV data of sample S1 by minimizing the squared sum of the relative errors $ \sigma = \sum_{V=V_{\text{min}}}^{V_{\text{max}}} \big( \big( I_{\text{S1,exp}} (V) - I_{\text{Tot}} (V) \big) / I_{\text{S1,exp}} (V) \big)^2$
%\begin{equation}
%   \sigma = \sum_{V=V_{\text{min}}}^{V_{\text{max}}} \Big( \big( I_{\text{S1,exp}} (V) - I_{\text{Tot}} (V) \big) / I_{\text{S1,exp}} (V) \Big)^2
%\end{equation}
between $V_{\text{min}} = \SI{0.39}{V}$ and $V_{\text{max}} = \SI{2.9}{V}$. Introducing an additional shunt conductance $G_P$ to the model (\ref{eq_LambertW}) did not improve the fit ($G_P \xrightarrow{} 0$) and is consequently omitted for further analysis.

To overcome the ambiguity of the determination of the contact areas $A_i$, the following procedure is proposed: Assuming a constant specific on resistance $R_{\text{on},i} A_i = \text{const.}$ for all current paths, the area $A_i$ can be found by $ A_i = (R_{\text{on},i}/R_{\text{on,Tot}}) \cdot A_{\text{Tot}}$,
%\begin{equation}
%    A_i = (R_{\text{on},i}/R_{\text{Tot}}) \cdot A_{\text{Tot}},
%\end{equation}
using the combination of parallel resistances $R_{\text{on,Tot}} = \left( \sum_{i=1}^N R_{\text{on},i}^{-1} \right)^{-1}$ with $A_{\text{Tot}} = \sum_{i=1}^N A_i$.

Fig. \ref{fig6} shows the experimental IV data of the diode on S1 and the model fit using (\ref{eq_LambertW}) with $N=\num{2}$. The main contribution to the current flow at higher voltages is characterized by a high ideality factor of $n_1=\num{1.43}$. Under the assumption $R_{\text{on},i} A_i = \text{const.}$, $A_1$ accounts for 99.994 \% of the total diode area, with a Schottky barrier of $\varphi_{B,1} = \SI{1.14}{eV}$. The current flow in the low voltage ($V<\SI{0.6}{V}$) region is governed by a small area patch ($A_2 = \SI{0.006}{\%} \times A_{\text{Tot}} \approx \SI{2}{\um^2}$) with a reduced effective Schottky barrier of $\varphi_{B,2} = \SI{0.89}{eV}$ and a slightly reduced ideality factor $n_2=\num{1.23}$. Several other diodes of S1, with diameters $d$ of both 100 µm and 200 µm, exhibit a similar IV characteristic regarding $n_1$, $n_2$, $\varphi_{B,1}$ and $\varphi_{B,2}$. Regardless of $d$ and the fitted total specific on-resistance $R_{\text{on,Tot}} A_{\text{Tot}}$, $A_2$ is found to be very similar for all these diodes and in the range $\SI{1}{\um^2} < A_2 < \SI{3}{\um^2}$. Due to the small number of different diameters, no clear area specific influence could be identified. Since the influence of horizontal current spreading is omitted in the previous analysis, the areas $A_2$ might be overestimated.

Despite the increased thickness of the i-layer of S1, the specific on-resistance $R_{\text{on,Tot}} A_{\text{Tot}} = \SI{350}{m\ohm.cm^2}$ of the diode presented in Fig. \ref{fig6} is comparable to the ones measured for the samples S2 and S3. For $V>\SI{2}{V}$, $R_{\text{on,Tot}} A_{\text{Tot}}$ is slightly underestimated by the fit as can be seen in the linear scaled inset in Fig. \ref{fig6}.

\begin{figure}[!t]
    \centerline{\includegraphics[width=\columnwidth]{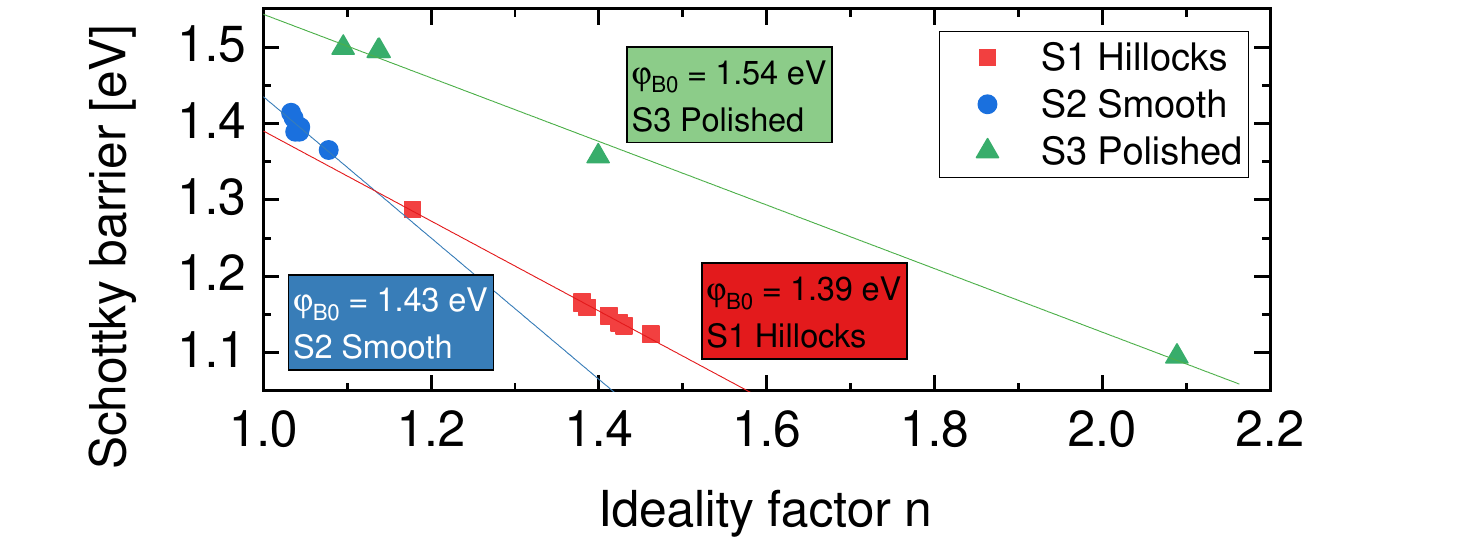}}
    \caption{Schottky barrier of various diodes from the three samples S1, S2 and S3. The solid lines are linear fits which yield $\varphi_B^0$}
    \label{fig7}
\end{figure}

Note that whereas the fitted barriers $\varphi_{B,i}$ drastically depend on the assumption made for the determination of the areas $A_i$ for the model fit, both $n_i$ and $R_{\text{on,Tot}} A_{\text{on}}$ are unaffected. Discarding $R_{\text{on},i} A_i = \text{const.}$ and assuming, for example, equal areas for all current paths (i.e. $A_i = A_{\text{Tot}}/N$), yields $\varphi_{B,1} \approx \varphi_{B,2} \approx \SI{1.13}{eV}$ with the same $n_i$, $R_{\text{on,Tot}} A_{\text{Tot}}$ and $\sigma$.

We also compared the results for $n$, $\varphi_B$ and $R_{\text{on}}A$ when fitting the IV data of the diodes of the samples S2 and S3 to both (\ref{eq_TEM}) and (\ref{eq_LambertW}). Both models yield very similar results with relative variations smaller than 2 \%. Consequently, the modified Lambert-W model (\ref{eq_LambertW}) constitutes a valid approach to characterize the Schottky contact properties of dual barrier diodes.

The Schottky barriers determined by IV are all smaller than those determined by CV. This trend is understandable when considering that an effective barrier $\varphi_{B\text{,eff}}$ governs the current transport during IV-measurement. Small patches with a reduced barrier height or a random distribution (e.g. Gauss like) of barrier heights reduce the effective barrier \cite{Werner.02.1991}\cite{Tung.11.2001} as measured by IV. As was discussed by Tung \textit{et al.} \cite{Tung.11.2001}, a Gauss-like distribution of barrier heights yields a dependence of the barrier heights $\varphi_{B\text{,eff}}(n) = \varphi_B^0 - 3(n-1) V_{bb} /2 $ on the ideality factor where $V_{bb}$ denotes the band bending. Fig. \ref{fig7} shows $\varphi_B (n)$ for several diodes of the three samples. For the dual barrier diodes of S1, $\varphi_{B,1}(n)$ is displayed. The $\varphi_B^0$ of S1 and S2 are comparable to the barrier heights obtained by CV, whereas $\varphi_B^0$ of the polished sample S3 is still notably smaller than the value measured by CV. This conduction mechanism might be attributed to a different surface modification due to the H\textsubscript{2} plasma treatment of S3 after polishing. A different treatment (e.g. longer plasma exposure or etching with H\textsubscript{2}/O\textsubscript{2} plasma) may clarify this observation.

\subsubsection{Reverse IV characteristics}

\begin{figure}[!t]
    \centerline{\includegraphics[width=\columnwidth]{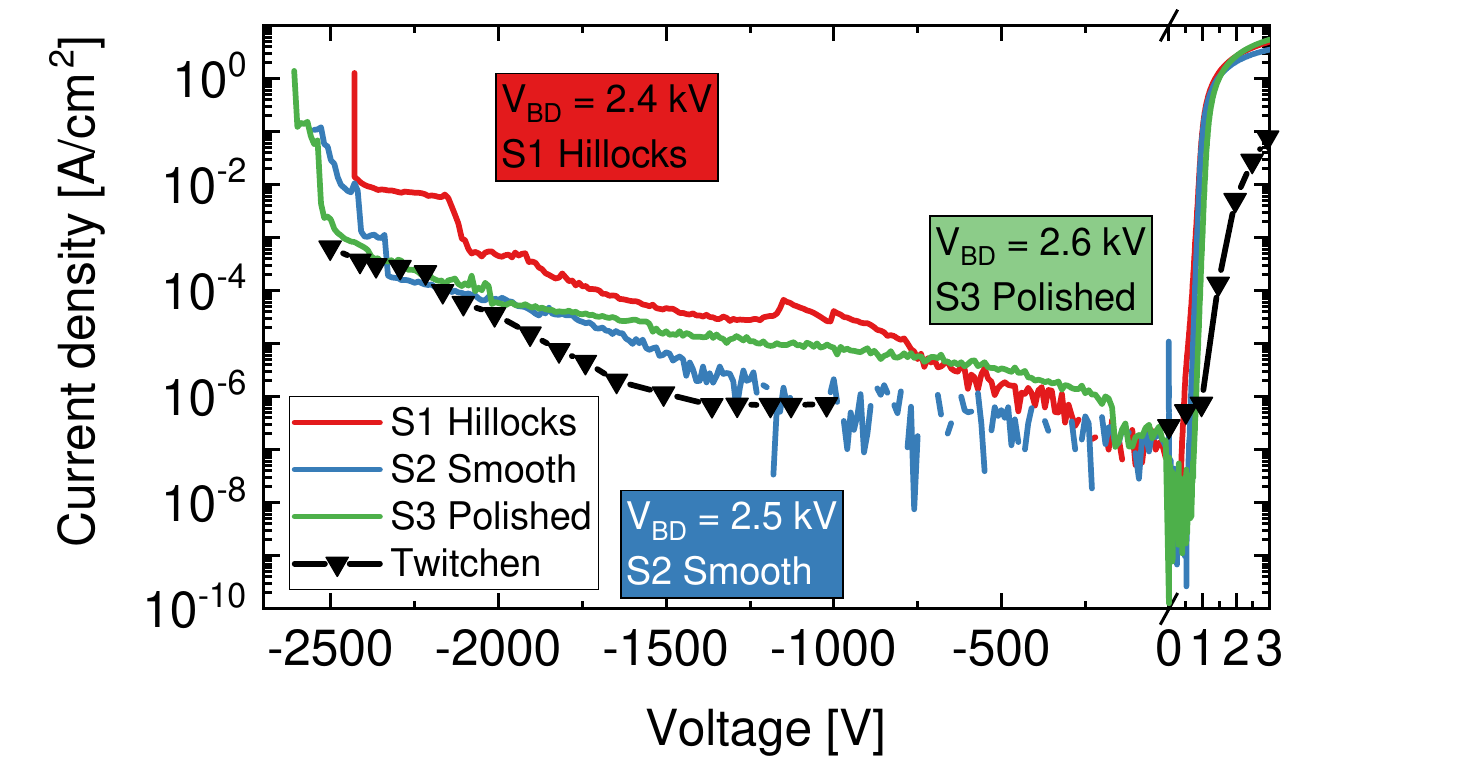}}
    \caption{Reverse IV curves of the samples S1, S2, and S3 in comparison with data from Twitchen \cite{Twitchen.05.2004}.}
    \label{fig8}
\end{figure}

The reverse IV characteristics of the diamond Schottky diodes are presented in Fig. \ref{fig8}. The data reported by Twitchen \textit{et al.} \cite{Twitchen.05.2004} is shown for comparison. The diode on the smooth as-grown sample S2 and the diode on the polished sample S3 exhibit a very low leakage current density $J_{\text{Rev}} < \SI{{}e-4}{A/cm^{-2}}$ for reverse voltages up to $\SI{2.2}{kV}$. An irreversible breakdown occurred at $V_{\text{BD}} = \SI{2.5}{kV}$ and $V_{\text{BD}} = \SI{2.6}{kV}$, respectively. The diode on the sample exhibiting a rougher surface with hillocks (Sample S1) shows a slightly higher leakage current over the entire voltage range. Breakdown occurs at $V_{\text{BD}} = \SI{2.4}{kV}$.

The breakdown occurred in the punch trough (PT) regime for all three diodes, since the theoretical epi-layer thickness required for the non-punch trough case \cite{Sze.2007} $W_{\text{NPT}} = \left( \frac{2 \epsilon_d V_{\text{BD}}}{q(N_A-N_D)} \right)^{1/2} > d_{\text{i-layer}}$. Considering the reciprocal punch through factor as defined by Chicot \textit{et al.} \cite{Chicot.10.2016} $\eta = d_{\text{i-layer}} / W_{\text{NPT}}$, the diodes in this paper have a PT factor of $\eta \approx \num{0.6}$ (S1) and $\eta \approx \num{0.2}$ (S2, S3), close to the optimal value of $\eta \approx \num{0.7}$ \cite{Chicot.10.2016}. With $\eta < \num{0.07}$, the factor of Twitchen’s diode is much lower which is induced by the low doping density $N_A - N_D < \SI{{}e13}{cm^{-3}}$ \cite{Twitchen.05.2004}. The low doping density results in a higher on-resistance and lower current density in forward direction.

When calculating the maximum electric field in a simplified, one-dimensional model using \cite{Sze.2007} $E_{\text{Max}} = \frac{V_{\text{BD}}}{d_{\text{i-layer}}} + \frac{q (N_A - N_D) d_{\text{i-layer}}}{2 \epsilon_d},$ sample S1 (Hillocks) and S3 (Polished) exhibit a similar breakdown field of $E_{\text{Max}} \approx \SI{1.2}{MV/cm}$, whereas sample S2 (Smooth) withstands a roughly 30 \% higher field with $E_{\text{Max}} \approx \SI{1.6}{MV/cm}$. The reduced breakdown field likely arises from the rough surface (S1) and the high defect density (S2). Due to a field enhancement caused by the inhomogeneous surface of S1, the local field at breakdown may even be higher than $E_{\text{Max}}$  as calculated by the simple 1D model. Since no edge termination was applied for all three diodes, a field enhancement is expected for all three devices \cite{Ikeda.02.2009}.

\begin{figure}[!t]
    \centerline{\includegraphics[width=\columnwidth]{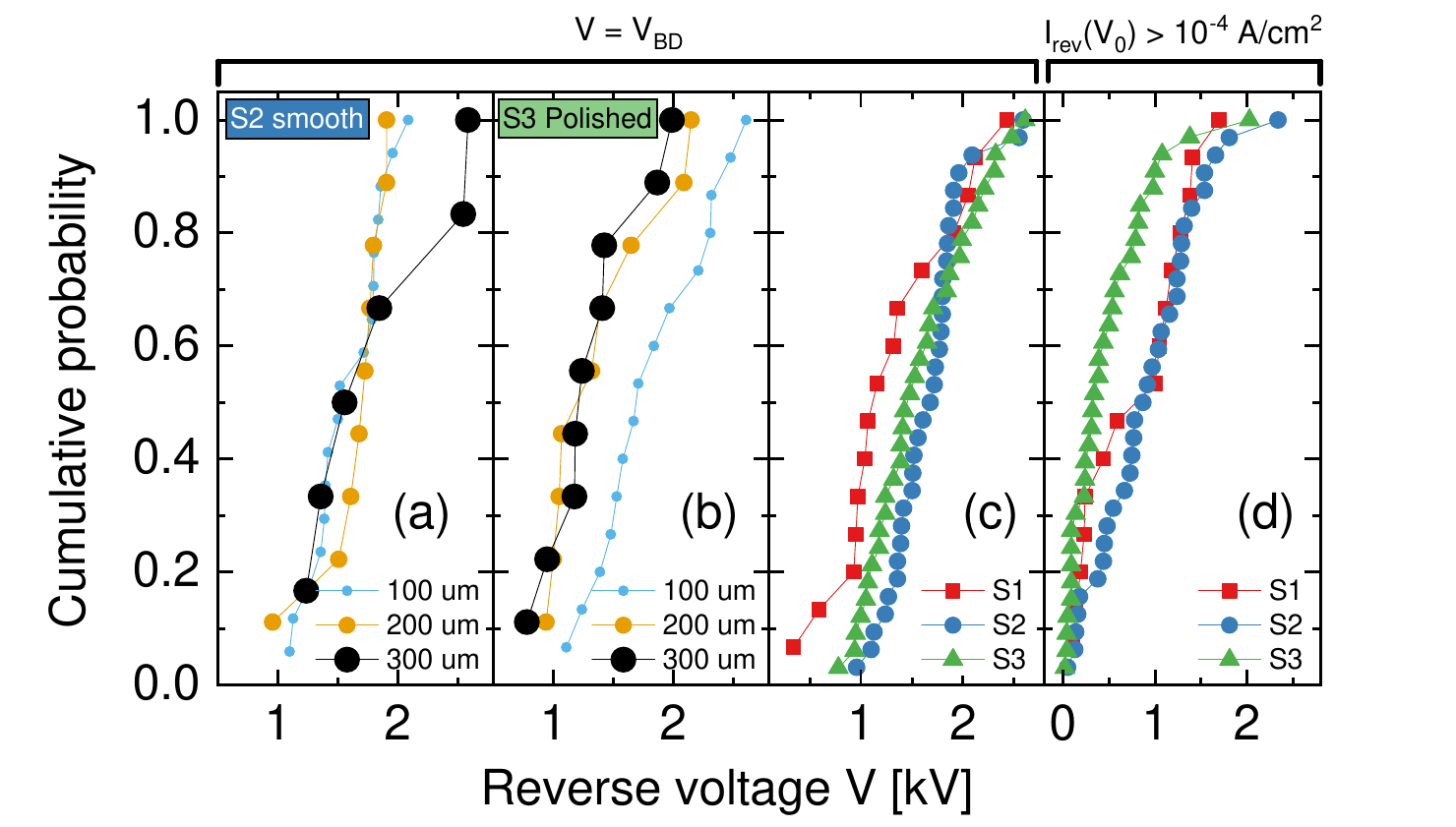}}
    \caption{(a), (b) and (c) cumulative probability of the breakdown voltage
$V=V_{\text{BD}}$ for S2, S3, and all samples, respectively. (d) cumulative probability of the reverse voltage $V=V_0$ satisfying $J_{\text{Rev}} > \SI{{}e-4}{A/cm^2}$ for all samples.}
    \label{fig9}
\end{figure}

A statistical analysis of the reverse IV characteristics of several diodes of each sample is shown in Fig. \ref{fig9}. The cumulative probability ($\text{CP}$) of the breakdown voltage $V_{\text{BD}}$  [i.e. $\text{CP} (V_{\text{BD}})$] of S2 shows no prominent dependence on the diode diameter [Fig. \ref{fig9} (a)]. The highest blocking diodes are statistical outliers. As can be seen in Fig. \ref{fig9} (b), $\text{CP} (V_{\text{BD}})$ of the small diodes on S3 is notably shifted to higher breakdown voltages. This implies a higher density of “killer defects” on the polished sample \cite{Schroder.2006}, limiting the area feasible for diode fabrication. The diameter-dependent $\text{CP} (V_{\text{BD}})$ of S1 behaves similar to $\text{CP} (V_{\text{BD}})$ of S2 and is therefore not shown.

The comparison of $\text{CP} (V_{\text{BD}})$ of all three samples in Fig. \ref{fig9} (c) shows a slightly steeper slope of $\text{CP} (V_{\text{BD}})$ for the smooth as grown sample S2 when compared to the polished sample S3, indicating a more homogeneous crystal quality. The rough sample S1 exhibits generally lower breakdown voltages, which may be attributed to local field enhancement caused by the surface roughness.

To analyze the reverse current, Fig. \ref{fig9} (d) plots $\text{CP} (V=V_0)$ with $J_{\text{Rev}} > \SI{{}e-4}{A/cm^2}$. Both of the untreated samples S1 and S2 show a similar behavior, while the polished sample S3 exhibits a pronounced increased probability for a higher reverse current density. The diameter dependent $\text{CP} (V=V_0)$ of S3 (not shown) is similar to $\text{CP} (V_{\text{BD}})$ in Fig. \ref{fig9} (b), indicating that the high defect density of the substrate is also responsible for the higher reverse current density.

No prominent relationship was established between the reverse IV characteristics and the Schottky barrier or the ideality factor.

To conclude the statistical analysis, polishing a sample could potentially allow the fabrication of high voltage blocking diodes. However, due to the higher reverse current density and the restricted feasible device area on low quality substrates, only substrates with a low defect density enable a reasonable device fabrication.

\begin{figure}[!t]
    \centerline{\includegraphics[width=\columnwidth]{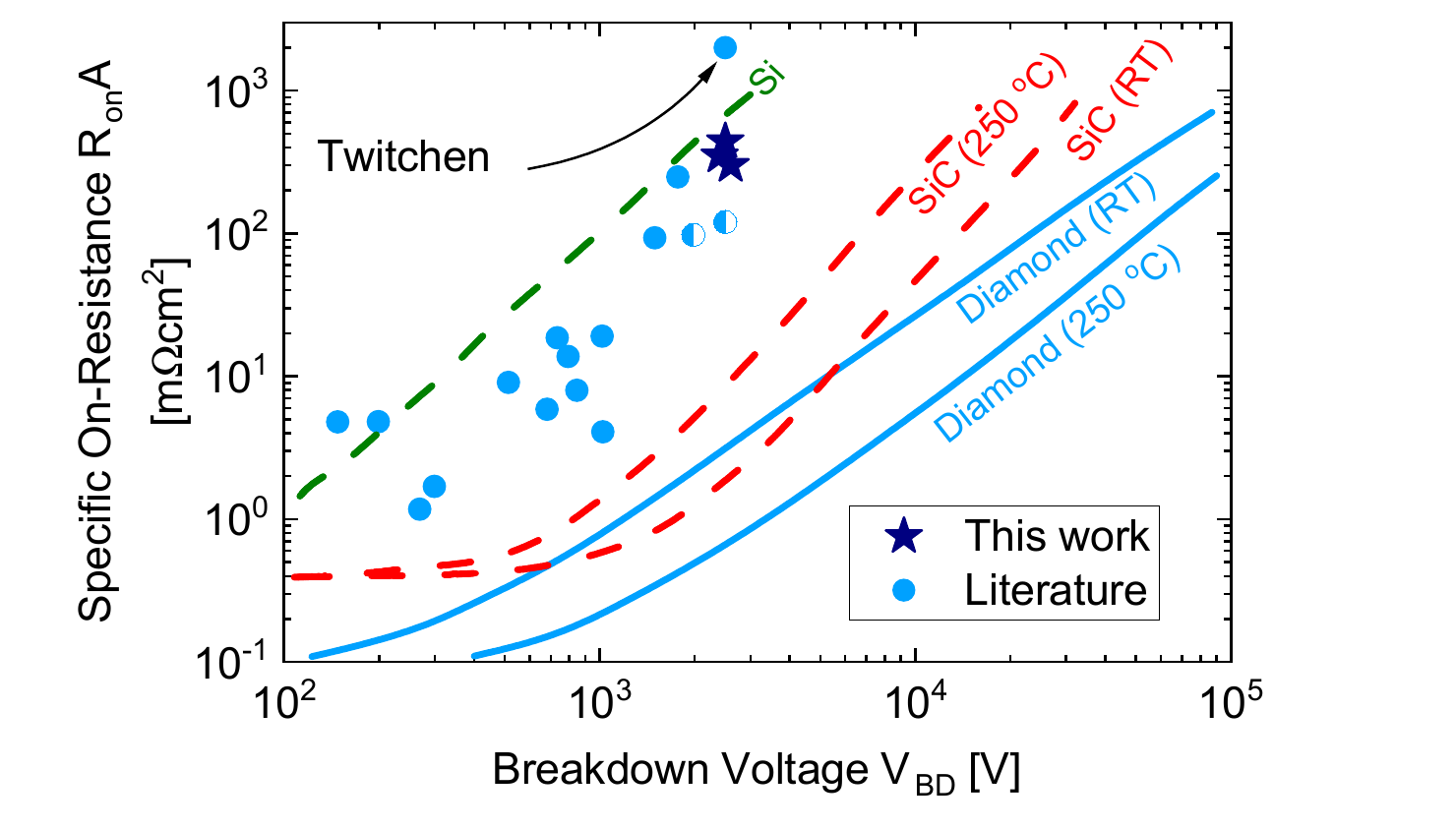}}
    \caption{Baliga plot showing the trade-off between $R_{\text{on}}A$ and $V_{\text{BD}}$ for unipolar diode devices on diamond, SiC an Si. Literature data shows experimental data of diamond diodes and is taken from \cite{Umezawa.05.2018}. The half-filled circles correspond to data points from \cite{Umezawa.05.2018} for which no related publication was found}
    \label{fig10}
\end{figure}

As a possibility to classify the quality of the present work, Fig. \ref{fig10} plots $R_{\text{on}}A$ versus $V_{\text{BD}}$ (“Baliga plot”) for various diamond diodes published to date \cite{Umezawa.05.2018}, with a comparison of the theoretical limits for unipolar power devices on Si, SiC, and diamond. Note that for some of the data points from \cite{Umezawa.05.2018}, no related publication could be found (half-filled circles). These seemingly “better” diodes (regarding the $\text{BFOM}$) are therefore omitted in the comparison in this work. The diodes presented in this work outperform the diode presented by Twitchen \cite{Twitchen.05.2004}, with a Baliga Figure of Merit $\text{BFOM}$ of $(\num{18}, \num{15}, \num{21}) \: \si{MW/cm^2}$ for the samples S1, S2, S3, signifying an approx. 7-time increase over Twitchen’s diode ($\SI{3}{MW/cm^2}$).

Table I summarizes the sample parameters and the data gathered from the IV and CV measurements. 

{\renewcommand{\arraystretch}{1.1}	%slightly larger rows
\begin{table}
\caption{Sample Details and Electrical Properties}
\setlength{\tabcolsep}{3pt}
\begin{tabular}{C{60pt}C{45pt}C{40pt}C{40pt}C{40pt}}
\hline
Quantity& 
Unit&
S1&
S2&
S3 \\
\hline
i-layer thickness&
$\si{um}$&
28&
17&
23 \\
%\hline
Surface quality&
-&
\textbf{Hillocks}&
\textbf{Smooth}&
\textbf{Polished} \\
%\hline
RMS roughness&
$\si{nm}$&
12 - 20&
$<$0.8&
1.5\\
%\hline
Schottky barrier \par (IV)&
$\si{eV}$&
(0.89),\par1.14&
1.38&
1.51\\
%\hline
Schottky barrier \par (CV)&
$\si{eV}$&
1.45&
1.45&
1.80\\
%\hline
Ideality factor&
-&
(1.23), \par 1.43&
1.03&
1.08\\
%\hline
$R_{\text{on}}A$&
$\si{m\ohm.cm^2}$&
350&
440&
300\\
%\hline
$V_{\text{BD}}$&
$\si{kV}$&
2.4&
2.5&
2.6\\
%\hline
1D $E_{\text{Max}}$&
$\si{MV/cm}$&
1.2&
1.6&
1.2\\
%\hline
BFOM&
$\si{MW/cm^2}$&
18&
15&
21\\
%\hline
PT Factor $\eta$&
-&
0.6&
0.2&
0.2 \\
\hline
\end{tabular}
\label{tab1}
\end{table}
{\renewcommand{\arraystretch}{}

\section{Summary}

In the present work, three homoepitaxial grown diamond layers with different surface morphologies were employed for the fabrication of high voltage Schottky diodes. Even though the diodes from all three samples were able to block at least $\SI{2.4}{kV}$, the diodes on both the polished and the hillock-rich sample (S3 and S1) showed a reduced performance with respect to the maximum breakdown field ($E_{\text{Max}} \approx \SI{1.2}{MV/cm}$). Presumably, this is attributed to the high defect density of the epitaxial layer (S3) and a field enhancement due to the rough surface (S1).

The diodes fabricated on the smooth as-grown sample S2, whose homoepitaxial layer was grown on a high quality diamond substrate, exhibited a breakdown field of $E_{\text{Max}} \approx \SI{1.6}{MV/cm}$. Similar to the polished sample S3, a low reverse current density $J_{\text{Rev}} < \SI{{}e-4}{A/cm^2}$ was measured for reverse voltages up to $\SI{2.2}{kV}$. The rectification ratio of S2 was as high as $10^8$, with an on-resistance of $R_{\text{on}}A = \SI{440}{m\ohm.cm^2}$. This is approx. 5 times lower than the on-resistance of diamond diodes with similar breakdown voltages \cite{Twitchen.05.2004}, yielding the Baliga Figure of Merit $\text{BFOM} = \SI{15}{MW/cm^2}$. Due to their lower on-resistance, both the diode on the polished and on the hillock-rich sample exhibited an even higher $\text{BFOM}$ of $\SI{21}{MW/cm^2}$ and $\SI{18}{MW/cm^2}$, respectively. Further work is required to assess whether the reduced series resistance of the polished diode is affected by the surface treatment.

A statistical analysis of several diodes of each sample showed that even though a low-quality substrate with a polished surface is suitable for high voltage blocking diodes, the high density of crystal defects both reduces the feasible device area and increases the reverse current density.

By employing the Lambert-W function, the current transport of dual barrier diodes was modelled with high accuracy. Future work may explore the influence of the diode’s diameter on the calculated patches of reduced barrier heights, examining any influence of local surface inhomogeneities on the current transport.

To conclude, we showed that the surface morphology and the crystal quality play an important role in the device performance of diamond Schottky diodes. We expect a further increase of the breakdown voltage if a suitable edge termination is applied.

\bibliographystyle{myIEEEtran}
\bibliography{ms}

% Generated by IEEEtran.bst, version: 1.13 (2008/09/30)
\begin{thebibliography}{10}
\providecommand{\url}[1]{#1}
\csname url@samestyle\endcsname
\providecommand{\newblock}{\relax}
\providecommand{\bibinfo}[2]{#2}
\providecommand{\BIBentrySTDinterwordspacing}{\spaceskip=0pt\relax}
\providecommand{\BIBentryALTinterwordstretchfactor}{4}
\providecommand{\BIBentryALTinterwordspacing}{\spaceskip=\fontdimen2\font plus
\BIBentryALTinterwordstretchfactor\fontdimen3\font minus
  \fontdimen4\font\relax}
\providecommand{\BIBforeignlanguage}[2]{{%
\expandafter\ifx\csname l@#1\endcsname\relax
\typeout{** WARNING: IEEEtran.bst: No hyphenation pattern has been}%
\typeout{** loaded for the language `#1'. Using the pattern for}%
\typeout{** the default language instead.}%
\else
\language=\csname l@#1\endcsname
\fi
#2}}
\providecommand{\BIBdecl}{\relax}
\BIBdecl

\bibitem{Nazare.2001}
M.~H. Nazar{\'e} and A.~J. Neves, Eds., \emph{Properties, growth and
  applications of diamond}.\hskip 1em plus 0.5em minus 0.4em\relax London:
  INSPEC, 2001, vol.~26. ISBN 0852967853

\bibitem{Isberg.09.2002}
J.~Isberg, J.~Hammersberg, E.~Johansson, T.~Wikstr{\"o}m, D.~J. Twitchen, A.~J.
  Whitehead, S.~E. Coe, and G.~A. Scarsbrook, ``High carrier mobility in
  single-crystal plasma-deposited diamond,'' \emph{Science}, vol. 297, no.
  5587, pp. 1670--1672, 09.2002. doi: 10.1126/science.1074374

\bibitem{Umezawa.05.2018}
H.~Umezawa, ``Recent advances in diamond power semiconductor devices,''
  \emph{Mater. Sci. Semicond. Process.}, vol.~78, pp. 147--156, 05.2018. doi:
  10.1016/j.mssp.2018.01.007

\bibitem{Volpe.08.2010}
P.-N. Volpe, P.~Muret, J.~Pernot, F.~Omn{\`e}s, T.~Teraji, F.~Jomard,
  D.~Planson, P.~Brosselard, N.~Dheilly, B.~Vergne, and S.~Scharnholtz, ``High
  breakdown voltage schottky diodes synthesized on p-type cvd diamond layer,''
  \emph{phys. stat. sol. (a)}, vol. 207, no.~9, pp. 2088--2092, 08.2010. doi:
  10.1002/pssa.201000055

\bibitem{Volpe.11.2010}
P.-N. Volpe, P.~Muret, J.~Pernot, F.~Omn{\`e}s, T.~Teraji, Y.~Koide, F.~Jomard,
  D.~Planson, P.~Brosselard, N.~Dheilly, B.~Vergne, and S.~Scharnholz,
  ``Extreme dielectric strength in boron doped homoepitaxial diamond,''
  \emph{Appl. Phys. Lett.}, vol.~97, no.~22, p. 223501, 11.2010. doi:
  10.1063/1.3520140

\bibitem{Bormashov.05.2017}
{Bormashov, V. S. et al.}, ``Thin large area vertical schottky barrier diamond
  diodes with low on-resistance made by ion-beam assisted lift-off technique,''
  \emph{Diamond Relat. Mater.}, vol.~75, pp. 78--84, 05.2017. doi:
  10.1016/j.diamond.2017.02.006

\bibitem{Teraji.06.2009}
T.~Teraji, Y.~Garino, Y.~Koide, and T.~Ito, ``Low-leakage p-type diamond
  schottky diodes prepared using vacuum ultraviolet light/ozone treatment,''
  \emph{J. Appl. Phys.}, vol. 105, no.~12, p. 126109, 06.2009. doi:
  10.1063/1.3153986

\bibitem{Umezawa.02.2007}
H.~Umezawa, T.~Saito, N.~Tokuda, M.~Ogura, S.-G. Ri, H.~Yoshikawa, and S.-i.
  Shikata, ``Leakage current analysis of diamond schottky barrier diode,''
  \emph{Appl. Phys. Lett.}, vol.~90, no.~7, p. 073506, 02.2007. doi:
  10.1063/1.2643374

\bibitem{Teraji.05.2012}
T.~Teraji, M.~Y. Liao, and Y.~Koide, ``Localized mid-gap-states limited reverse
  current of diamond schottky diodes,'' \emph{J. Appl. Phys.}, vol. 111,
  no.~10, p. 104503, 05.2012. doi: 10.1063/1.4712437

\bibitem{Twitchen.05.2004}
D.~J. Twitchen, A.~J. Whitehead, S.~E. Coe, J.~Isberg, J.~Hammersberg,
  T.~Wikstrom, and E.~Johansson, ``High-voltage single-crystal diamond
  diodes,'' \emph{IEEE Trans. Electron Devices}, vol.~51, no.~5, pp. 826--828,
  05.2004. doi: 10.1109/TED.2004.826867

\bibitem{Reiner.05.2019}
R.~Reiner, V.~Zuerbig, L.~Pinti, P.~Reinke, D.~Meder, S.~Moench, R.~Quay,
  V.~Cimalla, C.~E. Nebel, and O.~Ambacher, ``Diamond schottky-diode in a
  non-isolated buck converter,'' in \emph{PCIM Europe 2019}, 05.2019, pp.
  212--216.

\bibitem{Funer.03.1998}
M.~F{\"u}ner, C.~Wild, and P.~Koidl, ``Novel microwave plasma reactor for
  diamond synthesis,'' \emph{Appl. Phys. Lett.}, vol.~72, no.~10, pp.
  1149--1151, 03.1998. doi: 10.1063/1.120997

\bibitem{Widmann.04.2016}
C.~J. Widmann, W.~M{\"u}ller-Sebert, N.~Lang, and C.~E. Nebel, ``Homoepitaxial
  growth of single crystalline cvd-diamond,'' \emph{Diamond Relat. Mater.},
  vol.~64, pp. 1--7, 04.2016. doi: 10.1016/j.diamond.2015.12.016

\bibitem{Teraji.2015}
T.~Teraji, ``High-quality and high-purity homoepitaxial diamond (100) film
  growth under high oxygen concentration condition,'' \emph{J. Appl. Phys.},
  vol. 118, no.~11, p. 115304, 2015. doi: 10.1063/1.4929962

\bibitem{Tallaire.01.2008}
A.~Tallaire, M.~Kasu, K.~Ueda, and T.~Makimoto, ``Origin of growth defects in
  cvd diamond epitaxial films,'' \emph{Diamond Relat. Mater.}, vol.~17, no.~1,
  pp. 60--65, 01.2008. doi: 10.1016/j.diamond.2007.10.003

\bibitem{Ohmagari.09.2011}
S.~Ohmagari, T.~Teraji, and Y.~Koide, ``Non-destructive detection of killer
  defects of diamond schottky barrier diodes,'' \emph{J. Appl. Phys.}, vol.
  110, no.~5, p. 056105, 09.2011. doi: 10.1063/1.3626791

\bibitem{Akashi.06.2018}
N.~Akashi, A.~Seki, H.~Saito, F.~Kawai, and S.~Shikata, ``Influence of
  dislocations to the diamond sbd reverse characteristics,'' \emph{Mater. Sci.
  Forum}, vol. 924, pp. 212--216, 06.2018. doi:
  10.4028/www.scientific.net/MSF.924.212

\bibitem{Kato.08.2015}
Y.~Kato, H.~Umezawa, and S.-i. Shikata, ``X-ray topographic study of defect in
  p$-$ diamond layer of schottky barrier diode,'' \emph{Diamond Relat. Mater.},
  vol.~57, pp. 22--27, 08.2015. doi: 10.1016/j.diamond.2015.03.021

\bibitem{Sze.2007}
S.~M. Sze and K.~K. Ng, \emph{Physics of semiconductor devices}, 3rd~ed.\hskip
  1em plus 0.5em minus 0.4em\relax Hoboken, NJ: Wiley-Interscience, 2007. ISBN
  9780471143239

\bibitem{Nebel.2017}
C.~E. Nebel, ``General properties of diamond,'' in \emph{Nanodiamonds}.\hskip
  1em plus 0.5em minus 0.4em\relax Elsevier, 2017, pp. 1--24. ISBN
  9780323430296

\bibitem{Barjon.02.2011}
J.~Barjon, T.~Tillocher, N.~Habka, O.~Brinza, J.~Achard, R.~Issaoui, F.~Silva,
  C.~Mer, and P.~Bergonzo, ``Boron acceptor concentration in diamond from
  excitonic recombination intensities,'' \emph{Phys. Rev. B}, vol.~83, no.~7,
  02.2011. doi: 10.1103/PhysRevB.83.073201

\bibitem{Blank.08.2015}
{Blank, V. D. et al.}, ``Power high-voltage and fast response schottky barrier
  diamond diodes,'' \emph{Diamond Relat. Mater.}, vol.~57, pp. 32--36, 08.2015.
  doi: 10.1016/j.diamond.2015.01.005

\bibitem{Defives.03.1999}
D.~Defives, O.~Noblanc, C.~Dua, C.~Brylinski, M.~Barthula, V.~Aubry-Fortuna,
  and F.~Meyer, ``Barrier inhomogeneities and electrical characteristics of
  ti/4h-sic schottky rectifiers,'' \emph{IEEE Trans. Electron Devices},
  vol.~46, no.~3, pp. 449--455, 03.1999. doi: 10.1109/16.748861

\bibitem{Ma.01.2006}
X.~Ma, P.~Sadagopan, and T.~S. Sudarshan, ``Investigation on barrier
  inhomogeneities in 4h-sic schottky rectifiers,'' \emph{phys. stat. sol. (a)},
  vol. 203, no.~3, pp. 643--650, 01.2006. doi: 10.1002/pssa.200521017

\bibitem{Umezawa.09.2009}
H.~Umezawa, K.~Ikeda, N.~Tatsumi, K.~Ramanujam, and S.-i. Shikata, ``Device
  scaling of pseudo-vertical diamond power schottky barrier diodes,''
  \emph{Diamond Relat. Mater.}, vol.~18, no.~9, pp. 1196--1199, 09.2009. doi:
  10.1016/j.diamond.2009.04.013

\bibitem{Banwell.02.2000}
T.~C. Banwell and A.~Jayakumar, ``Exact analytical solution for current flow
  through diode with series resistance,'' \emph{Electron. Lett.}, vol.~36,
  no.~4, p. 291, 02.2000. doi: 10.1049/el:20000301

\bibitem{Jung.11.2009}
W.~Jung and M.~Guziewicz, ``Schottky diode parameters extraction using lambert
  w function,'' \emph{Mater. Sci. Eng., B}, vol. 165, no. 1-2, pp. 57--59,
  11.2009. doi: 10.1016/j.mseb.2009.02.013

\bibitem{Olikh.07.2015}
O.~Y. Olikh, ``Review and test of methods for determination of the schottky
  diode parameters,'' \emph{J. Appl. Phys.}, vol. 118, no.~2, p. 024502,
  07.2015. doi: 10.1063/1.4926420

\bibitem{OrtizConde.06.2012}
A.~Ortiz-Conde, D.~Lugo-Mu{\~n}oz, and F.~J. Garc{\'i}a-S{\'a}nchez, ``An
  explicit multiexponential model as an alternative to traditional solar cell
  models with series and shunt resistances,'' \emph{IEEE J. Photovoltaics},
  vol.~2, no.~3, pp. 261--268, 06.2012. doi: 10.1109/JPHOTOV.2012.2190265

\bibitem{Corless.12.1996}
R.~M. Corless, G.~H. Gonnet, D.~E.~G. Hare, D.~J. Jeffrey, and D.~E. Knuth,
  ``On the lambertw function,'' \emph{Adv. Comput. Math.}, vol.~5, no.~1, pp.
  329--359, 12.1996. doi: 10.1007/BF02124750

\bibitem{Werner.02.1991}
J.~H. Werner and H.~H. G{\"u}ttler, ``Barrier inhomogeneities at schottky
  contacts,'' \emph{J. Appl. Phys.}, vol.~69, no.~3, pp. 1522--1533, 02.1991.
  doi: 10.1063/1.347243

\bibitem{Tung.11.2001}
R.~T. Tung, ``Recent advances in schottky barrier concepts,'' \emph{Mater. Sci.
  Eng., R}, vol.~35, no. 1-3, pp. 1--138, 11.2001. doi:
  10.1016/S0927-796X(01)00037-7

\bibitem{Chicot.10.2016}
G.~Chicot, D.~Eon, and N.~Rouger, ``Optimal drift region for diamond power
  devices,'' \emph{Diamond Relat. Mater.}, vol.~69, pp. 68--73, 10.2016. doi:
  10.1016/j.diamond.2016.07.006

\bibitem{Ikeda.02.2009}
K.~Ikeda, H.~Umezawa, N.~Tatsumi, K.~Ramanujam, and S.-i. Shikata,
  ``Fabrication of a field plate structure for diamond schottky barrier
  diodes,'' \emph{Diamond Relat. Mater.}, vol.~18, no. 2-3, pp. 292--295,
  02.2009. doi: 10.1016/j.diamond.2008.10.021

\bibitem{Schroder.2006}
D.~K. Schroder, \emph{Semiconductor material and device characterization},
  3rd~ed.\hskip 1em plus 0.5em minus 0.4em\relax Hoboken N.J.: {IEEE Press} and
  Wiley, 2006. ISBN 9780471739067

\end{thebibliography}

\end{document}